\documentstyle[prd,aps]{revtex}
\begin{document}

\title{Large quantum gravity effects: backreaction on matter}

\author{Rodolfo Gambini
\\{\em Instituto de F\'{\i}sica, Facultad de
Ciencias,\\Tristan Narvaja 1674, Montevideo, Uruguay}}

\author{Jorge Pullin\\  {\em Center for Gravitational Physics and Geometry}\\
{\em Department of Physics, 104 Davey Lab,}\\ {\em The Pennsylvania State 
University,}\\
{\em University Park, PA 16802}}

\maketitle
\vspace{-6cm} 
\begin{flushright}
\baselineskip=15pt
CGPG-97/4-1  \\
gr-qc/9703088\\
\end{flushright}
\vspace{5cm}

\begin{abstract}
We reexamine the large quantum gravity effects discovered by Ashtekar
in the context of $2+1$ dimensional gravity coupled to matter. We
study an alternative one-parameter family of coherent states of the
theory in which the large quantum gravity effects on the metric can be
diminished, at the expense of losing coherence in the matter
sector. Which set of states is the one that occurs in nature will
determine if the large quantum gravity effects are actually observable
as wild fluctuations of the metric or rapid loss of coherence of
matter fields.
\end{abstract}

${}$

One normally thinks that classical general relativity has a rather wide and
well defined domain of applicability as a realistic theory of nature. 
This belief is based on the expectation that if one were to consider a full 
quantum treatment of the theory, one could find states that minimize 
uncertainties in all the basic fields and that obey macroscopically 
the classical Einstein equations  of evolution. One expects that this picture
would break down only when very high energies come into play.

Ashtekar \cite{As} recently reexamined these beliefs by considering a
model of quantum gravity coupled to matter in $2+1$ dimensions. The
specific kind of matter is not very relevant to the discussion, but
for concreteness it is taken to be a scalar field. The theory is
completely integrable classically.  The metric is given by,
\begin{equation}
ds^2=e^{G \Gamma(r,t)} (t,r)(-dt^2+dr^2) +r^2 d\theta^2
\end{equation}
where $\Gamma$ is completely determined by the scalar field,
\begin{equation}
\Gamma(r,t)={1 \over 2} \int_0^r dr' r' [\dot{\phi}^2 +{\phi'}^2].
\end{equation}

So one can picture the theory as that of a scalar field living in a
fiducial flat spacetime that completely determines the metric through
the above expressions. One can quantize the model by considering a
Fock representation for the scalar field and thinking of the metric as
a quantum operator.  What Ashtekar observes is that if one considers a
coherent state for the scalar field and computes the uncertainties in
the metric one gets, in the high frequency regime (the frequency
refers to the coherent state of the scalar field),
\begin{equation}
\left(\Delta g_{rr}\over <g_{rr}>\right)^2 
\sim e^{{\cal N} (e^{2 G \hbar \omega})}
\end{equation}
where ${\cal N}$ is the occupation number of the coherent state. The
surprise here is that even if one considers a configuration with such
a low classical field amplitude that it corresponds to a ``one
photon'' (${\cal N}=1$) state, one finds that the uncertainty in the
metric is huge if the frequency of the photon is high. This is quite
unexpected, ie, one expected that the classical theory would break
down for large curvature effects, not that a field of low amplitude and
high frequency could 
have such dramatic effects on the metric. The effect is largely due to
the nonlinearity of gravity, which in turn implies the exponential
dependence of metric and field. It is not present, for instance, for
scalar electrodynamics \cite{Aspers}. The above effect is not
necessarily an artifact of $2+1$ dimensions, the above calculations
also represent $3+1$ gravity with one constant-norm Killing vector
field. Similar effects could also be present in other, more realistic
$3+1$ contexts \cite{Aspers}.

The issue we wish to raise in this note is up to what extent should
one consider states for this model exactly as the ones considered
above. Such states were precisely coherent states for the field, and
therefore are clearly a good choice to represent a classical state of
matter. In the quantization, since one thinks of the metric as a
derived operator, one does not pay too much attention to its
uncertainties while constructing the coherent states. But what if one
did? That is, what happens if we consider states that not only
minimize the uncertainties in the scalar field variables but also
simultaneously minimizes the uncertainties of the metric. Could one
avoid the large quantum gravity effects? Or would the price of
suppressing the large metric fluctuations translate in a rapid loss of
coherence of the matter fields? A quantitative answer to these
questions is the main purpose of this note.

For simplicity we will limit the discussion to one frequency mode, the
extension to many modes is immediate. We seek for states $|\psi>$ that
minimize\footnote{Our conventions are $[a,a^\dagger]=1$,
$[\hat{x},\hat{p}]=i$.},

\begin{equation}
{\cal F} = {(\Delta x)^2 +(\Delta p)^2 \over <x>^2+<p>^2} + \kappa
 \left(\Delta g
\over <g>\right)^2 -\beta (<\psi|\psi> -1) + \lambda 
[<x>^2 +<p>^2 - 2 |\alpha|^2].
\end{equation}
where $\hat{x} = (a+a^\dagger)/\sqrt{2}$ and $\hat{p} = i
(a^\dagger-a)/\sqrt{2}$. That is, we are seeking to minimize
uncertainties in $x$, $p$ and $g$ simultaneously. The parameter
$\kappa$ determines the relative weight of the uncertainties in the
metric with respect to the field variables in the minimization
process. The Lagrange multipliers $\lambda$ and $\beta$ are introduced
to require that the states are normalized and that the expectation
values of position and momenta have the correct scaling with the
energy, $|\alpha|$. This latter condition is equivalent to imposing
$\alpha =(<x>+i <p>)/\sqrt{2}$. If one takes $\kappa=0$ the states
that minimize the above functional are the ordinary coherent states,
\begin{equation}
|\alpha> = \sum_{n=0}^\infty e^{-{|\alpha|^2\over 2}} 
{\alpha^n\over \sqrt{n!}}
|n>
\end{equation}

This suggests to consider as proposal for states that will minimize the above
functional,
\begin{equation}
|\psi> = \sum_{n=0}^\infty b_n |n>
\end{equation}
where $|n>$ are the $n$-photon states of frequency $\omega$ for the 
scalar field in $2+1$ dimensions, see \cite{As}.

In terms of the above states, we get,
\begin{eqnarray}
<\psi|\psi> &=& \sum_{n=0}^\infty b_n^* b_n =1,\label{1}\\
{\sqrt{2}\over 2} <x+ip> &=& \sum_{m=0}^\infty \sqrt{m+1}\, b_m^*
b_{m+1} = \alpha, \label{2}\\
(\Delta x)^2 +(\Delta p)^2 &=& <x^2> +<p^2> -2 |\alpha|^2 = 
\sum_{m=0}^\infty (2 m +1) b^*_m b_m -2 |\alpha|^2,\\
<g> &=& \sum_{m=0}^\infty e^{m \Omega} b_m^* b_m\equiv g,\\
<g^2> &=& \sum_{m=0}^\infty e^{2 m \Omega} b_m^* b_m \equiv C,
\end{eqnarray}
where the metric operator was defined as in \cite{As} $\hat{g} =
 \exp(\Omega a^\dagger a)$, $\Omega$ being the ratio of the frequency
 $\omega$ to the Planck frequency.

With the above relations we can proceed to minimize ${\cal F}$. We get 
as minimization condition 
\begin{equation}
(2m +1) b^*_m + 2 \kappa |\alpha|^2 (g^{-2} e^{2\Omega m} -
2 C g^{-3} e^{\Omega m}) b^*_m-\beta b_m^* - \lambda ( 2 \sqrt{m} b^*_{m-1}
\alpha^* + 2 \sqrt{m+1} b^*_{m+1} \alpha) = 0\label{3}
\end{equation}
where we have redefined $\beta$ and $\lambda$ to absorb a factor
$2|\alpha|^2$. Given a value of $\Omega$, the solution depends on
$\kappa$ and $\alpha$. The joint solution of the above equation
(\ref{3}) with the constraint equations (\ref{1},\ref{2}) imposed by
the Lagrange multipliers completely determines the problem, giving
values for $b_m$, $\beta$ and $\lambda$.

We have been unable to solve the above minimization condition in closed
form. We will perform a perturbative analysis in terms of $\Omega$, which
is a small parameter (it is the ratio of the frequency of the state to the
Planck frequency) in the classical, sub-Planckian domain. Before
performing the perturbative analysis, we introduce new rescaled
variables, $\ell = \lambda^{-1}$ and 
\begin{equation}
b_m = {\alpha^m e^{-{|\alpha|^2\over 2}} \ell^m \over \sqrt{m}!} B_m.
\end{equation}

In terms of these we can rewrite the minimization condition as 
\begin{equation}
2 m (B^*_m - \ell^{-2} B^*_{m-1}) +2 |\alpha|^2 (B_m^* - B^*_{m+1}) - 
\beta B^*_m + 2 \kappa |\alpha|^2 (g^{-2} e^{2\Omega m}-2 C g^{-3}
e^{\Omega m}) B^*_m = 0.\label{3p}
\end{equation}

Let us start by studying the case $\Omega=0$, which we will consider
the zeroth order in the perturbative analysis we wish to perform. The
solution to the joint set of equations (\ref{1},\ref{2},\ref{3p}) is 
\begin{equation}
B_m =1,\qquad \ell =1 \qquad \beta = -2 \kappa |\alpha|^2,
\end{equation}
and leads to the usual coherent states. This is equivalent to solving
for $\kappa=0$ and $\Omega$ arbitrary. For the usual coherent states
$(\Delta x)^2 + (\Delta p)^2 =1$.

We now study higher order perturbations in $\Omega$. It is
straightforward to see that the term  in ${\cal F}$ dependent on the
metric does not give contributions to first order in $\Omega$, so to
see a nontrivial effect of the new proposal we need to go to second
order. To compute to second order we write $B_m = 1 +\Omega^2
B^{(2)}_m$, $\ell=1+\Omega^2 \ell^{(2)}$ and $\beta = 
-2 \kappa |\alpha|^2 + \beta^{(2)}$, and also,
\begin{eqnarray}
<g> &=& 1 + \Omega \sum_{m=0}^\infty m b^*_m b_m + {\Omega^2\over 2} 
\sum_{m=0}^\infty m^2 b^*_m b_m,\\
b_m &=& {\alpha^m e^{-{|\alpha|^2\over 2}}\over \sqrt{m}!} + O(\Omega^2).
\end{eqnarray}
Now, noting that 
\begin{eqnarray}
\sum_{m=0}^\infty {m |\alpha|^{2m}\over m!} e^{-{|\alpha|^2}} = 
e^{-|\alpha|^2} |\alpha|^2 {\partial \over \partial |\alpha|^2}
e^{|\alpha|^2} &=& |\alpha|^2,\\
\sum_{m=0}^\infty {m^2 |\alpha|^{2m}\over m!} e^{-|\alpha|^2} = 
e^{-|\alpha|^2} |\alpha|^2 {\partial \over \partial |\alpha|^2}
|\alpha|^2 {\partial \over \partial |\alpha|^2}
e^{|\alpha|^2} &=& |\alpha|^4+|\alpha|^2,
\end{eqnarray}
we get,
\begin{eqnarray}
<g> &=& 1 + |\alpha|^2 \Omega + {|\alpha|^4 \Omega^2\over 2} +
{|\alpha|^2 \Omega^2 \over 2} +O(\Omega^3),\\
<g^2> &=& 1 + 2 |\alpha|^2 \Omega + 2{|\alpha|^4 \Omega^2} +
2 {|\alpha|^2 \Omega^2 } +O(\Omega^3),
\end{eqnarray}
and from here the recursion relation (\ref{3p}) reads,
\begin{equation}
2 m (B_m^{(2)} -B_{m-1}^{(2)}) + 2 |\alpha|^2 (B_m^{(2)} -
B_{m+1}^{(2)}) -\beta^{(2)} + 4 m \ell^{(2)} + 2 \kappa |\alpha|^2
(m^2 - 2 m |\alpha|^2+|\alpha|^4) =0.
\end{equation}

One immediately sees that the $B^{(2)}_m$'s only depend on $\alpha$
through $|\alpha|^2$, since $g$ and $C$ are real. This has important 
implications, it means that the states will not disperse in time,
that is, under time evolution,
\begin{equation}
b_m(\alpha,\kappa) \longrightarrow b_m(\alpha e^{i \Omega t}, \kappa),
\end{equation}
which in turn implies that $<x>$ and $<p>$ satisfy the classical 
equations of motion, ie, the states have all the required properties of 
coherent states \cite{Kl}.

The solution of the recursion relation is,
\begin{eqnarray}
\ell &=& 1 -\Omega^2 \kappa |\alpha|^2,\\
\beta&=& -2 \kappa |\alpha|^2 -4 \kappa |\alpha|^4 \Omega^2,\\
b_m &=& {\alpha^m \over \sqrt{m}!} e^{-{|\alpha|^2 \over 2}} \left(
1 - {\kappa |\alpha|^6 \Omega^2 \over 2} + m \Omega^2 (\kappa
|\alpha|^4 + {\kappa |\alpha|^2 \over 2})-{\kappa |\alpha|^2 \over 2}
m^2 \Omega^2\right).
\end{eqnarray}

An analogous calculation yields the $O(\Omega^3)$ corrections,
\begin{eqnarray}
\ell^{(3)} &=& - \kappa |\alpha|^2,\\
b_m^{(3)} &=& {\alpha^m \over \sqrt{m}!} e^{-{|\alpha|^2 \over 2}} \left[
{2 \over 3} \kappa |\alpha|^8 + {5 \over 2} \kappa
|\alpha|^6 + m (-\kappa |\alpha|^6 + 2 \kappa |\alpha|^4 + {5 \over 6}
\kappa |\alpha|^2) \right.\nonumber\\
&&\left.+ m^2 ( \kappa |\alpha|^4 - {\kappa |\alpha|^2
\over 2}) - {m^3 \kappa |\alpha|^2\over 3}]\right]
\end{eqnarray}

From here we can compute the uncertainties in the quantities of
physical interest,
\begin{eqnarray}
(\Delta x)^2 + (\Delta p)^2 &=& 1 + \Omega^3 \left({2\over 3} \kappa
|\alpha|^{10} + 8 \kappa |\alpha|^8\right),\\
\left(\Delta g \over g\right)^2 &=& 
\left(\Delta g \over g\right)^2_{\kappa =0} - 2 \kappa |\alpha|^6
\Omega^4 - \Omega^5 \kappa({2 \over 3} |\alpha|^{12} +{22\over 3}
|\alpha|^{10} - 8 |\alpha|^8 + 6 |\alpha|^6).
\end{eqnarray}
So we see that as we increase the parameter $\kappa$ the fluctuations
in the matter fields increase and the fluctuations in the
gravitational field decrease. This allows the large quantum gravity
effects to manifest themselves in ways different from Ashtekar's
original proposal, ie, they may arise as early losses of coherence of
the matter fields, long before the gravitational perturbations are
significant. Notice also the rather high powers of $\alpha$ involved,
which suggests that these kinds of effects might be more detectable
than the ones due to ordinary coherent states. 

The above results imply the existence of a one-parameter family of
coherent states for the gravitational field coupled to
matter. However, because we have pursued a perturbative power series
approach, one may question if the results presented are rigorous, in
the sense that the power series discussed may fail to converge. To
show that one really has a family of states of physical interest, we
need to discuss the convergence of the series discussed.

Therefore, we need to study the behavior of the coefficients for large
values of $m$. We get,
\begin{equation}
b_m = {e^{-{|\alpha|^2 \over 2}} |\alpha|^m \over \sqrt{m}!} \left(
1 -{\Omega^2 \kappa |\alpha|^2 m^2 \over 2} - {\Omega^3
\kappa |\alpha|^2 m^3 \over 3} -\ldots - 
{\Omega^p (2^p -2) \over p p!}\kappa |\alpha|^2 m^p - \ldots\right),
\end{equation}
so,
\begin{equation}
|b_m| <{e^{-{|\alpha|^2 \over 2}} |\alpha|^m \over \sqrt{m}!} \left(1 +
\sum_{p=1}^\infty {\Omega^p 2^p m^p \over p!} \kappa |\alpha|^2\right)
={e^{-{|\alpha|^2 \over 2}} |\alpha|^m \over \sqrt{m}!} \left[
\left(e^{2m\Omega} -1\right) \kappa |\alpha|^2 +1 \right]
\end{equation}
and therefore the series $\sum_{m} b^*_m b_m$ converges,
\begin{equation}
\sum b_m^* b_m < (1 -\kappa |\alpha|^2) + \kappa |\alpha|^2
e^{|\alpha|^2(e^{2\Omega}-1)}.
\end{equation}

Summarizing, we see that in the context of the $2+1$ example analyzed
by Ashtekar \cite{As} the gravitational field coupled to matter
admits a more general one-parameter family of coherent states. These
are states that minimize the combined uncertainties of the matter
fields and the gravitational field and are preserved under evolution.
The ``large quantum gravity effects'' pointed out by Ashtekar are
present for all these states. That is, if one has matter modes of high
frequency, fluctuations in the physical quantities may become large,
even if one has ``one photon'' of high frequency. However, depending
on the value of the parameter $\kappa$ that characterizes our family,
the fluctuations can be concentrated on the metric (as in Ashtekar's
original example) or in the matter fields. In this sense the ``large
quantum gravity effects'' may not be necessarily manifested as in
Ashtekar's original suggestion as large fluctuations in the metric,
but as early losses of coherence of the matter fields. This also opens
the possibility for the effects to become observable at lower values
of the frequency. 

A natural question is, which are the ``correct'' states to describe
nature from within this one-parameter family. This will require a more
detailed analysis of the situation in question, and how the material
system under study interacts with its environment. That will
ultimately determine which of these states are relevant and therefore
in which range of physical parameters can one expect to see the large
effects. One may think that these only manifest themselves for
frequencies that are comparable with Planck's frequency, and therefore
are mostly of academic interest. However, it is worthwhile pointing
out that they not only depend on $\Omega$ but also on the value of
$|\alpha|^2$, which can be very large in realistic systems. Indeed,
our perturbative calculations break down for $\Omega\sim 10^{-25}$ if
$|\alpha|^2 \sim 10^{20}$. In actual high power lasers, one can get
$\Omega\sim 10^{-28}$ with $|\alpha|^2 \sim 10^{15}$ so the predicted
decoherence is not necessarily  that far from observability.

An intriguing possibility is if these effects might be present in the
collapse of neutron stars. The outer core of neutron stars is in a
superfluid state, which could be regarded as a coherent state of
matter. The problem here is that the frequency in question is largely
determined by the rotation rate of the star (the frequency is related
with the kinetic energy of the matter). These rates are very low
compared with the Planck frequency. However, when the star collapses
and the outer regions become trapped within the horizon, local
velocity gradients become very high. It is questionable however, if
these local velocity gradients are the relevant ones for
quantization. The detailed discussion of this situation clearly
exceeds the scope and possibilities of the $2+1$ example we have been
discussing all along and application of the current results can only
be considered at present as speculative. However, the possibility that
large quantum gravity effects could imply a loss of coherence of
matter or large fluctuations of the metric in neutron stars and
therefore may affect the collapse scenario is attractive enough to
merit further investigation of these kinds of effects.

We wish to thank Abhay Ashtekar and Curt Cutler for discussions. 
This work was also supported in part by grants
NSF-INT-9406269, NSF-PHY-9423950, by funds of the
Pennsylvania State University and its Office for Minority Faculty
Development, and the Eberly Family Research Fund at Penn State.  We also
acknowledge support of CONICYT and PEDCIBA (Uruguay).  JP also
acknowledges support from the Alfred P.  Sloan Foundation through an
Alfred P.  Sloan fellowship.


\begin{references}
\bibitem{As} A. Ashtekar, Phys. Rev. Lett. {\bf 77}, 4864 (1996).
\bibitem{Aspers} A. Ashtekar, private communication.
\bibitem{Kl} J. Klauder, ``Coherent states'', World Scientific, Singapore,
(1985).
\end{references}
\end{document}